# Introduction of DC line structures into a superconducting microwave 3D cavity

Wei-Cheng Kong[1,2], Guang-Wei Deng[1,2], Shu-Xiao Li[1,2], Hai-Ou Li[1,2], Gang Cao[1,2], Ming Xiao[1,2], and Guo-Ping Guo[1,2,+]

1 Key Laboratory of Quantum Information, University of Science and Technology of China, Chinese Academy of Sciences, Hefei 230026, China

2 Synergetic Innovation Center of Quantum Information and Quantum Physics, University of Science and Technology of China, Hefei, Anhui 230026, China

+ Correspondence to: gpguo@ustc.edu.cn

We report a technique that can noninvasively add multiple DC wires into a 3D superconducting microwave cavity for electronic devices that require DC electrical terminals. We studied the influence of our DC lines on the cavity performance systematically. We found that the quality factor of the cavity is reduced if any of the components of the electrical wires cross the cavity equipotential planes. Using this technique, we were able to incorporate a quantum dot (QD) device into a 3D cavity. We then controlled and measured the QD transport signal using the DC lines. We have also studied the heating effects of the QD by the microwave photons in the cavity.

Hybrid structures that can combine different quantum devices and a three-dimensional waveguide cavity[1–4] (3D cavity) are currently undergoing rapid development. This type of cavity has multiple advantages over the two-dimensional coplanar waveguide (CPW)-type cavity[5,6] that has been used in previous circuit quantum electrodynamics (QED) experiments. First, this type of cavity has a much larger mode volume, and is much less sensitive to surface dielectric losses.[1] Second, the devices inside the cavity are surrounded by a well-controlled electromagnetic environment, which limits the probability of relaxation through spontaneous emission into multiple modes.[1,7] Additionally, the point of maximum electric field intensity is easy to locate, and is usually at the center, corresponding to the cavity working frequency at the lowest resonant mode, and is much easier to calculate than that of CPW cavities. In coupling to Josephson junction qubits,[1] flux qubits[4,8] and electron spin ensembles,[3] this structure has already been used to reach strong coupling regions and manufacture squeezed photon states.[2,9,10] Recently, entanglement of two superconducting qubits that were spatially separated in two 3D cavities through a coaxial cable was observed.[11] By acting as both an entanglement bus and a readout circuit, the 3D cavity is promising for application to more complex multiple-qubit coupling structures. However, all the experiments above relied on only two ports (one for input and another for output) of the 3D cavity and the applied magnetic field.[12] Because of a lack of DC-biased control gates,[13,14] these cavities are incapable of more delicate control. Here, we introduce on-circuit DC lines to the 3D cavity to provide more precise gate control, and discuss the practical effects of these lines.

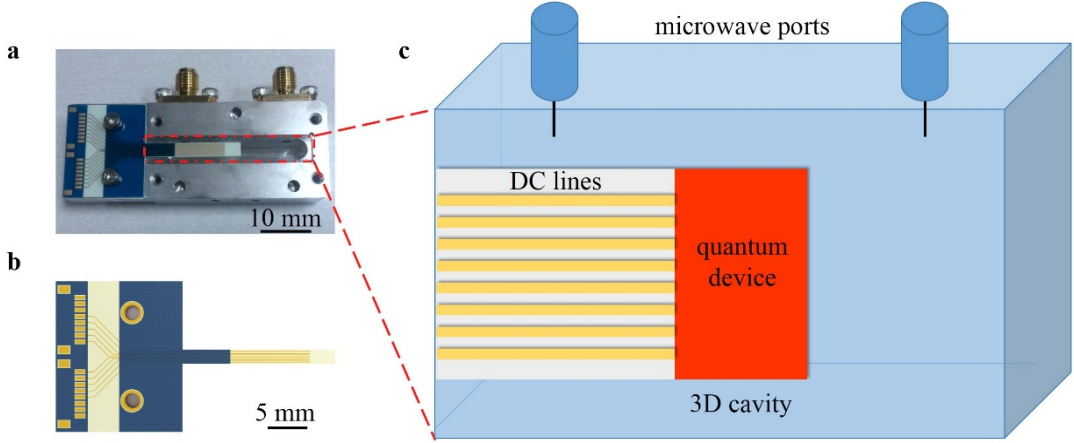

FIG. 1. DC line structure applied to the 3D cavity. (a) Half of our modified 3D cavity structure, with a platform on one side face, and a gap nearby to introduce the on-circuit DC lines. (b) Layout of the on-circuit DC line design. The circuit board was designed using Altium Designer Release 10 software and then fabricated using RO4350B sheet material from Rogers Corporation. Only the patterned areas are covered by copper (brown in the picture). (c) Schematic diagram of half of the hybrid device. To provide a simplified and artistic rendering, the model diagram does not reflect the real size.

The conventional 3D cavity structure is machined from aluminum, with a hollowed section in the middle. The external dimensions are 45.6 mm×25 mm×25.6 mm, and the hollow has dimensions of 30.6 mm×5 mm×17.8 mm. The three lowest resonant modes are the TE101, TE102 and TE103, with theoretical frequencies at approximately 9.50 GHz, 12.10 GHz and 15.54 GHz, respectively. We initially open a gap on one side face in the cross-section, as shown in Fig. 1(a). One of our custom-designed circuit boards is then fixed on the platform next to the gap. The circuit board reaches directly through the gap to the center of the cavity. Several copper DC lines are distributed in the same direction, with the layout as shown in Fig. 1(b). We fabricated a series of such circuit boards with various numbers of DC lines up to a maximum of sixteen lines, meaning that various quantum devices could be controlled using these DC lines. The blank bottom area of the circuit board is designed to hold the quantum devices. The complete scheme is shown in Fig. 1(c), where the DC lines and the quantum devices are in the same plane. Aluminum wires are bonded from the on-circuit DC lines to electrodes on the sample. In simulations using High Frequency Structure Simulator (HFSS) software, the gap causes a 200 kHz shift in our working mode (which is also the lowest resonant frequency) of $f_{TE101} = 9.502$ GHz, and a 5 kHz broadening of the corresponding bandwidth of about 200 kHz when compared with conventional modes,[1] which all have negligible values. This comparison is only in terms of the cavity design, and the DC lines and the "quantum device" are not included.

The directions of the microwave input/output ports are set on the *y*-axis, as shown in Fig. 2(a). Because of the symmetry of the electric field, the equipotential planes lie in the XZ plane. All on-circuit DC lines are distributed in the XY plane, parallel to the plane of the two ports. Theoretical calculations prove that the components of the DC lines in the directions of both the *x*- and *z*-axes have an insignificant effect on the cavity, if we also take the direction of the *z*-axis into consideration. However, the components along the *y*-axis cause serious damage to the cavity's working mode. Using HFSS software, these results have been simulated and are shown in Fig. 2(b). The four peaks reveal quality factors of the same order but different electromagnetic fields, particularly the fields on the XY plane where the DC lines and the quantum devices are distributed (as shown in Fig. 2(c)–(f)). We therefore simulate the corresponding electromagnetic fields at each resonant frequency, as shown in Fig. 2(c)–(f). Figure 2(d) and 2(f) prove that after application of a DC line along the *x*-axis or the *z*-axis, the TE101 working mode will still exist. However, when one DC line is applied along the *y*-axis, the TE102 mode appears at $f_{TE102} = 10.806$ GHz, while the corresponding TE101 mode disappears. We choose fields at points A and B for comparison. A is the maximum point of the field in Fig. 2(d), and B is the maximum point of the field in Fig. 2(e). These points are chosen to be near the locations of the original maximum field intensities of the TE101 mode and the TE102 mode. The numerical values are shown in Fig. 2(c)–(f). We confirm that the DC line components along the *x*-axis and the *z*-axis will strength the field, particularly around the center, while the component along the *y*-axis will weaken the field by several dozen times. Similarly disappointing results always occur as long as the DC lines have a component along the *y*-axis. Also, we found that the DC line positions do not affect the field too much, and only cause disturbances around the lines themselves. Ultimately, setting all of the DC lines in equipotential planes becomes the only choice. The final design is shown in Fig. 1(a).

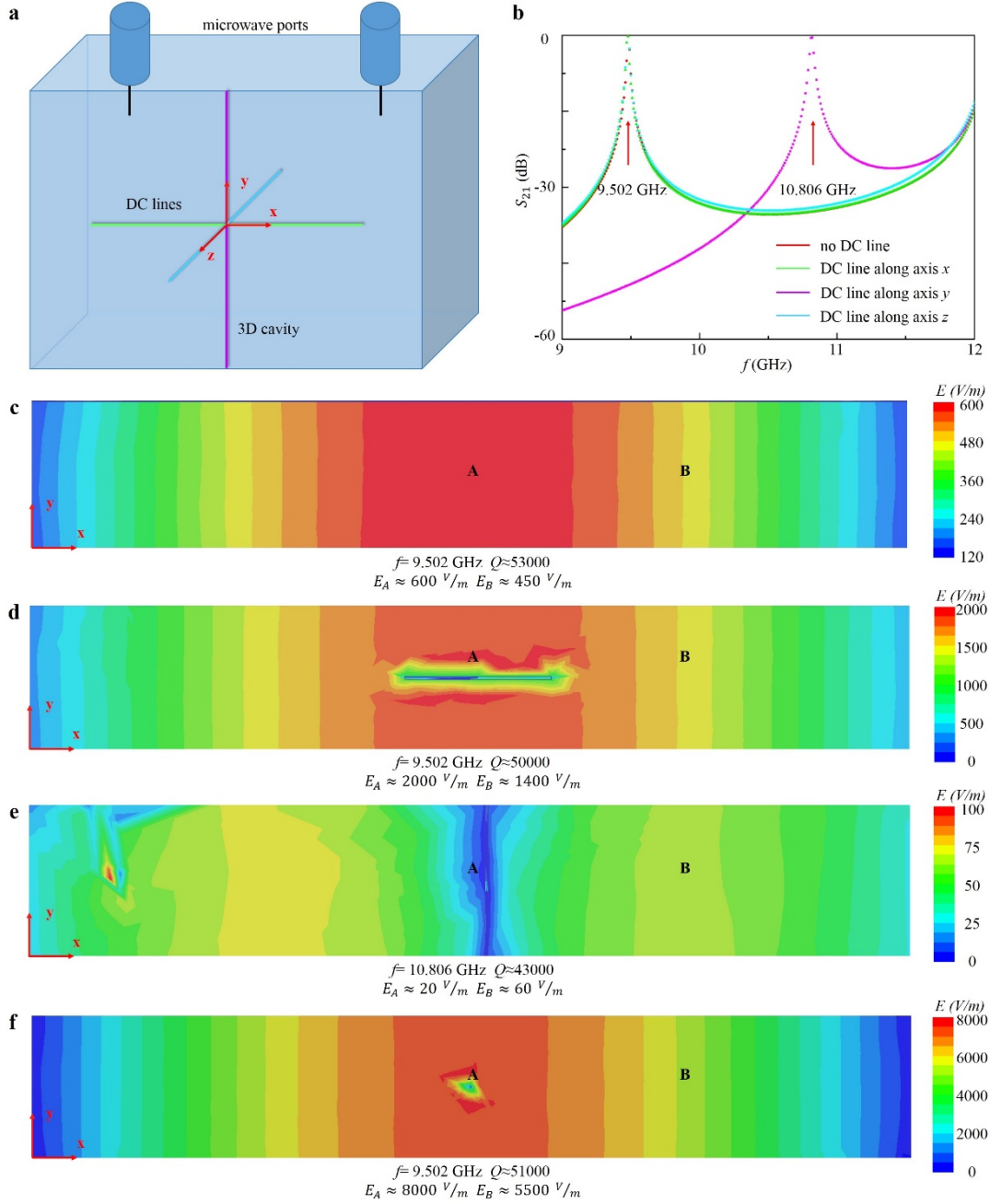

FIG. 2. (a) Schematic diagram of the directions of the DC lines in the 3D cavity. (b) Simulated transmission coefficient $S_{21}$ with DC lines across the central port in different directions. The DC lines in the equipotential planes (i.e., along the *x*-axis or the *z*-axis) will not change the lowest resonant mode of $f_{TE101} = 9.502$ GHz. When the DC lines are applied outside the equipotential plane (i.e., on the *y*-axis), the TE101 mode is destroyed, while the TE102 mode appears at 10.806 GHz. The electric field intensity distributions in the XY plane at the resonant frequency are shown in (c), (d), (e) and (f), and correspond to the situations where (c) no DC lines are applied, and where one DC line is applied along (d) the *x*-axis, (e) the *y*-axis, and (f) the *z*-axis. Points A and B are located near the maximum field intensities of the two lowest resonant modes (the TE101 mode and the TE102 mode, respectively), and the field intensities at these two points

are given under each picture. The scope of the electric field represents the real size of the 3D cavity.

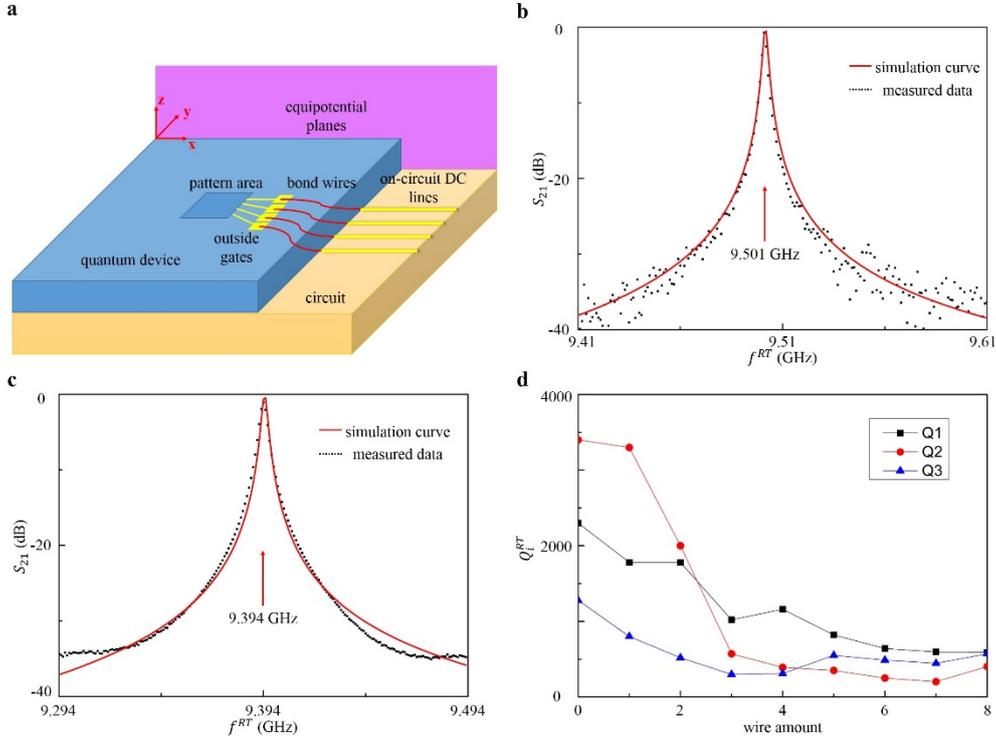

FIG. 3. Bonding wire testing at room temperature. (a) DC wiring architecture. The coordinate system used here is the same as that in Fig. 2(b). The quality factor of the cavity will decrease as long as these DC wires are partly out of the equipotential plane (XZ plane), i.e., they have a component on the *y*-axis. Only four groups of DC lines are drawn in the figure, while we can actually fabricate dozens of groups of lines. (b) Cavity signal when nothing additional applied to the 3D cavity; $f^{RT} = 9.501$ GHz and $Q^{RT} \approx 4300$. (c) Cavity signal when only applied with eight on-circuit DC lines; $f^{RT} = 9.394$ GHz and $Q^{RT} \approx 2300$. (d) Dependence of the quality factor $Q_i^{RT}$ on the number of bond wires when everything shown in (a) applied to the cavity. The subscripts 1, 2 and 3 correspond to the three lowest resonant modes at $f_{TE101} = 9.394$ GHz, $f_{TE102} = 11.977$ GHz and $f_{TE103} = 15.434$ GHz, respectively.

We used an intrinsic GaAs sample to test our DC lines at room temperature. The sample consisted of eight external gates connected to the pattern area shown in Fig. 3(a), unless the "pattern area" in the experiment was empty. Comparison of Fig. 3(b) and (c) shows that the influence of our circuit's DC line structure on our 3D cavity is insignificant; both signals were measured at room temperature using a network analyzer from Agilent Technology (E5071C ENA Series Network Analyzer). The resonant peak of a cavity without an on-circuit DC line structure is shown in (b), and (c) shows the corresponding result for a cavity with an eight-line circuit structure. The quality factor is damped from about 4300 to 2300, and these results are repeatable. The circuit causes a 110 MHz reduction in the resonant frequency, which was predicted by simulations. The data in Fig. 3(b) corresponds well with the simulation results, but the data in Fig.

3(c) do not match the simulation curve very well; we believe that this difference mainly stems from the technological limitations of the circuit fabrication process.

We used aluminum bond wires to connect our quantum device to the on-circuit DC lines, while the wire actually constitutes another type of DC line structure in our system. Figure 3(d) shows the effect of the number of bond wires on the cavity quality factor at room temperature. The circuit used here is the same circuit that was used in Fig. 3(c). The wirings have a critical impact on the quality factor. As Fig. 3(d) shows, the eight bond wires reduce $Q$ for the lowest resonant mode from 2300 to 600. When more than 12 wires are bonded to the sample, the resonant peak would be submerged in noise (not shown in Fig. 3). The problem was found to be the wire bonding process. The bond wires are ductile, leading to a complex and non-straight outline that produces wire components that are vertical to the equipotential planes (XZ plane in Fig. 3(a)) and have complex effects on the electromagnetic field. The magnitude of the field is weakened, and the quality factor decreases. With careful bonding, the effects of the wires can be reduced. With careful design of the circuit and the DC line distribution, the total impact of the wires on the cavity would be further reduced.

To test the tunability and practicality of our design, a gate-defined double-quantum-dot (DQD)/3D-cavity hybrid device was used as an example. Our DQD sample was sequentially fabricated by electron beam lithography and electronic beam evaporation on a GaAs/AlGaAs wafer,[5,15] and was measured below 10 mK using the Triton from Oxford Instruments (Triton 400 Cryogen-Free Dilution Refrigerator). The layout of the control gates is shown in Fig. 4(a), with eight electrodes, corresponding to the number of bond wires and on-circuit DC lines shown above. The transport signals were measured through the DC line structures. We located the Coulomb blockade regions, as shown in Fig. 4(b). The cavity resonant frequency is 9.348 GHz, with a total quality factor of 1350, as shown in Fig. 4(c). We could not read the quantum dot's signal through the cavity because of the weak coupling between the microwave photon and the quantum dot. We found that the dominant effect of the 3D cavity on our DC line-controlled DQD sample is the heating effect,[6] while the reaction of the cavity to the DQD was the reduction of the resonant frequency. A detailed analysis of the heating effect is shown in Fig. 4(d), where we vary the voltage of the left plunge gate, $V_2$, while the other gate voltages remain unchanged. The transport current and the cavity power are in positive correlation. The data in Fig. 4(e) were extracted from Fig. 4(d), when $V_2 = 0$ V and $V_4 = 0.95$ V. The data follow an exponential fit, which is the intrinsic characteristic of the heating effect.

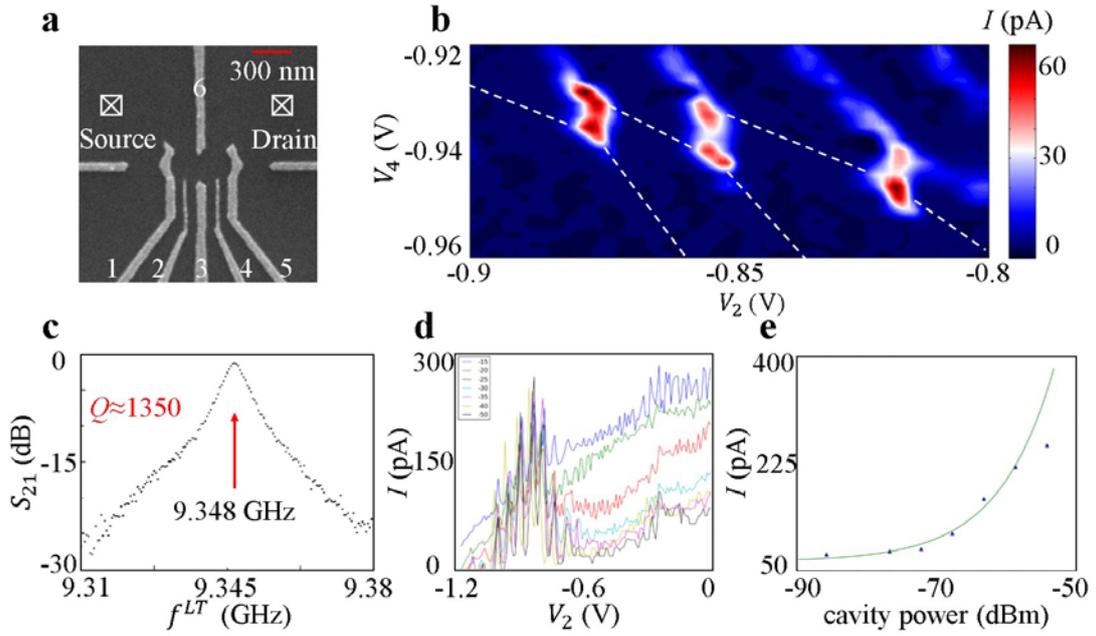

FIG. 4. (a) Scanning electron microscope image of gate-defined GaAs/AlGaAs DQD sample applied to our 3D cavity. All gates (six split gates, 1–6, and two ohmic contacts, Source and Drain) were connected to the measurement equipment through our DC line structures, as shown in Fig. 3(d). (b) Charge stability diagram for DQD, measured using the transport current through Source and Drain. (c) Cavity signal at low frequency. $f^{LT} = 9.348$ GHz and $Q^{LT} \approx 1350$. (d) Transport current through quantum dot as a function of $V_2$, under various microwave powers (from −90 dBm to −55 dBm) applied to the resonator, when $V_4 = 0.95$ V. (e) Transport current through quantum dot as a function of microwave power applied to the 3D cavity, when $V_2 = 0$ V and $V_4 = 0.95$ V. The green curve is an exponential fit.

In conclusion, we have designed a method to apply DC lines in a microwave 3D cavity, and have proved that the method is practicable. The DC line direction required to minimize the dissipation is clarified and the line number dependence of the cavity is reported. A hybrid DQD/cavity device was studied as an example, and showed tunability and potential for applications to future circuit QED structures and other experiments. We believe that our design will provide suitable options for establishment of entangled hybrid quantum systems.

We thank Professor Hong-Wen Jiang for helpful conversations, and Miao-Lei Zhang for experimental assistance. This work was supported by the National Fundamental Research Program (Grant No. 2011CBA00200), the National Natural Science Foundation (Grant Nos. 11222438, 11174267, 61306150, 11304301, 11274294, and 91121014), and the Chinese Academy of Sciences.